\documentclass[10pt]{article}
\usepackage{babel}
\usepackage{graphicx}
\usepackage{color}
\usepackage{bm}
\usepackage{longtable}
\usepackage{amsmath} 
\usepackage{amsfonts}
\usepackage{amssymb}
\usepackage{hyperref} 
\usepackage{mathtools}
\usepackage{subcaption}
\usepackage{cite} 
\usepackage[top=72pt,right=72pt,bottom=72pt,left=72pt]{geometry}
\begin{document}
\begin{center}
{\Large \bf Probabilistic Quantum Teleportation via 3-Qubit Non-Maximally Entangled GHZ State by Repeated Generalized Measurements
}\vskip 0.4cm
{\large Shamiya Javed\footnote{javedshamiya@allduniv.ac.in}, Ravi Kamal Pandey\footnote{ravikamalpandey@gmail.com}, Phool Singh Yadav\footnote{phsyadav@rediffmail.com}, Ranjana Prakash and Hari Prakash}
\vskip 0.4cm
Physics Department, University of Allahabad, Allahabad, India
\end{center}

\begin{abstract}
We propose a scheme of repeated generalized Bell state measurement (GBSM) for probabilistic quantum teleportation of single qubit state of a particle (say, 0) using 3-qubit non-maximally entangled (NME) GHZ state as a quantum channel. Alice keeps two qubits (say, 1 and 2) of the 3-qubit resource and the third qubit (say, 3) goes to Bob. Initially, Alice performs GBSM on qubits 0 and 1 which may lead to either success or failure. On obtaining success, Alice performs projective measurement on qubit 2 in the eigen basis of $\sigma_{x}$. Both these measurement outcomes are communicated to Bob classically, which helps him to perform a suitable unitary transformation on qubit 3 to recover the information state. On the other hand, if failure is obtained, the next attempt of GBSM is performed on qubits 0 and 2. This process of repeating GBSM on alternate pair of qubits may continue until perfect teleportation with unit fidelity is achieved. We have obtained analytical expressions for success probability up to three repetitions of GBSM. The success probability is shown to be a polynomial function of bipartite concurrence of the NME resource. The variation of success probability with the bipartite concurrence has been plotted which shows the convergence of success probability to unity with GBSM repetitions.  
\end{abstract}
\section{Introduction}\label{sec1}
Bennett et al. \cite{PhysRevLett.70.1895} proposed the scheme for quantum teleportation (QT) of quantum state of a two-level system (one qubit of information) from a sender, Alice, to a distant receiver, Bob, using a 2-qubit entangled state shared between Alice and Bob and a 2-bit classical communication channel. This idea has been demonstrated experimentally by Bouwmeester et al. \cite{bouwmeester1997experimental} using photon entangled pair. Prakash et al. \cite{prakash2007generalized} suggested that for QT of a single qubit information state at least 2-qubit entangled state resource is required. However, 3-qubit entangled state can also be used for this task. Karlsson et al. \cite{karlsson1998quantum} used a 3-qubit maximally entangled (ME) GHZ state to teleport a single qubit from sender, Alice to the receiver, Bob with the consent of a third party, Charlie, without whom cooperation, the information can not be teleported. This scheme is called controlled quantum teleportation (CQT).

Since, a ME state interacting with the environment tends to decohere \cite{prakash2008effect} and reduce to non-maximally entangled (NME) state \cite{Prakash2012}, the study of QT with NME resource have practical importance. Using ME resource (concurrence\cite{wootters1998entanglement}, $C=1$) QT is perfect with unit fidelity and unit success probability. However, for  NME resource ($C<1$), fidelity $F<1$ is obtained ($F=(2+C)/3$)\cite{Prakash2012}. Agrawal and Pati \cite{AGRAWAL200212,1464-4266-6-8-034} proposed a scheme of QT using NME resource with unit fidelity but less than unit success probability, which is termed as probabilistic  quantum teleportation (PQT). Wan Li Li et al. \cite{PhysRevA.61.034301} studied PQT using a generalized Bell basis and suggested that for maximum success probability there should be entanglement matching between entangled resource and the generalized Bell basis. In order to increase the success probability of PQT using NME resource, ample theoretical studies has been proposed by many groups of authors \cite{shi2006teleportation,Yan2010,Nie2009,xu2010tripartite,Ting_2005}.

Shi Biao \cite{shi2006teleportation} and Yang et al. \cite{Yan2010} considered QT with a 3-qubit NME GHZ state as a resource and used an ancilla along with a controlled rotation operation that ultimately results in PQT. The authors have suggested that if information $a|{0}\rangle+b|{1}\rangle$  is teleported as $\sim a\alpha|{0}\rangle+b\beta|{1}\rangle$ (with $\alpha$,$\beta$ real, and say,  $\alpha<\beta$), controlled rotation represented by unitary operator
	\begin{equation} U=|{0}\rangle\langle{0}|(|{0}\rangle\langle{0}|+|{1}\rangle\langle{1}|)+|{1}\rangle\langle{1}|[\cos{\eta}(|{0}\rangle\langle{0}|+|{1}\rangle\langle{1}|)+\sin{\eta}(|{0}\rangle\langle{1}|-|{1}\rangle\langle{0}|)]\label{eq1}
	\end{equation} where, $\cos{\eta}=\alpha/\beta$,   
 on the teleported particle (say 1), and an ancilla (say 2) initially in state $|{0}\rangle$  results in the state $\sim \alpha(a|{0}\rangle+b|{1}\rangle)_{1}|{0}\rangle_{2}+\sqrt{\beta^{2}-\alpha^{2}}b|{1}\rangle_{1}|{1}\rangle_{2}$ resulting in PQT. Nie et al. \cite{Nie2009} gave a scheme of controlled PQT using a four qubit cluster state.  In this scheme Alice performs BSM and the result is communicated to the controller, Charlie, who then performs GBSM on his pair of particles. With the help of GBSM outcomes of Alice and Charlie, Bob performs an appropriate composite unitary transformation on an auxiliary qubit and his qubit to recreate the information state. Probabilistic CQT using cluster type state has been proposed by Feng Xu et al. \cite{xu2010tripartite}, where both the sender and the controller perform BSM on their respective particles and communicate the measurement outcomes to a receiver. The receiver introduces an auxiliary particle and makes appropriate unitary operations to recover the original state with certain probability. Gao Ting et al. \cite{Ting_2005} proposed a scheme of CQT using an entangled state other than the GHZ state, in which first Charlie measures his particles in $\{|{0}\rangle,|{1}\rangle\}$ basis and sends the measurement result to Alice and Bob through some classical channel. Alice then performs BSM on her two particles (the unknown particle and shared entangled particle) and sends the measurement outcomes to Bob, who performs a unitary operation (depending on both Alice's and Charlie's measurement results) on his particle and retrieves the original state.
 
The QT utilising W state \cite{shi2002teleportation,joo2003quantum,cao2004probabilistic} has also been possible with a success probability upto 2/3. It has also been investigated using entangled coherent states \cite{PhysRevA.64.022313,prakash2007improving,prakash2008effect,MISHRA2015462,
 Pandey2019,prakash2019controlled}. 
Mishra and Prakash \cite{MISHRA2015462} devised an approach that used entangled coherent states to achieve perfect QT with as high success as required by performing repeated BSM. Recently, Pandey et al. proposed a scheme of obtaining high success standard quantum teleportation of information encoded in phase opposite coherent states by using resonant two-level atoms in a cavity \cite{pandey2021high}. Javed et al. \cite{javed2021high} proposed a scheme of repeated GBSM on the collapsed state to transmit a single qubit using a 2-qubit NME state as a resource. One of the drawback in this scheme is that after performing GBSM, Alice needs to send the particle to Bob allowing him to perform required unitary operation on this qubit to obtain the information state. The transmission of qubit over a quantum channel not only introduces noise to the information state but it also makes the information transfer delayed. In order to circumvent these issues our present scheme uses a 3-qubit NME GHZ state for QT of a single qubit from Alice to Bob, where the entangled state is priorly shared between Alice and Bob.

\begin{figure}[h!]
\centering
\includegraphics[scale=0.55]{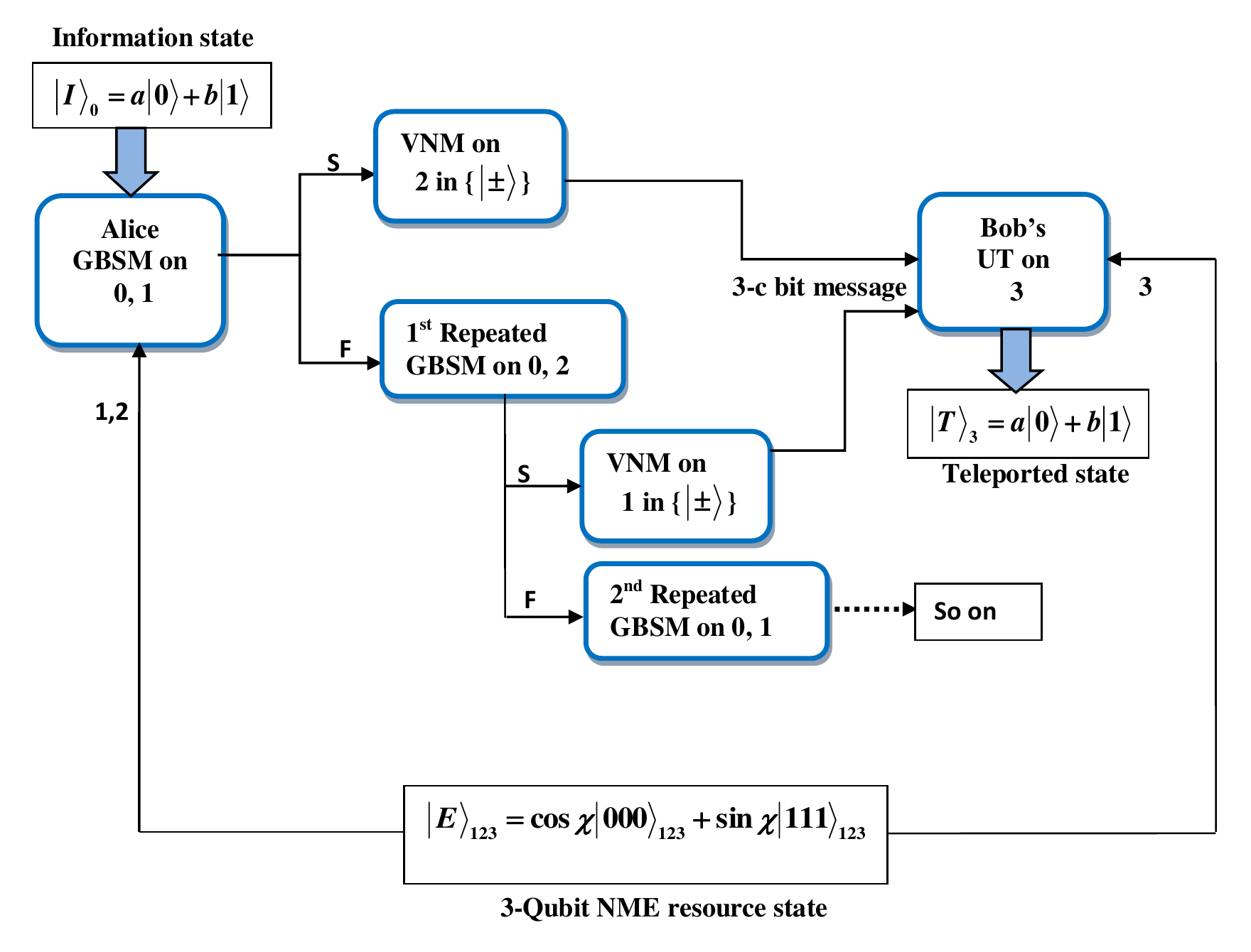}
\caption{\textit{Schematic diagram of the proposed scheme for PQT of a single qubit using 3-qubit NME GHZ state $|E\rangle_{123}$. Alice has information qubit (0) and two qubits (1,2) of the 3-qubit NME resource state. Alice makes generalized Bell state measurement (GBSM) on particles 0 and 1. On indicating success she makes von Neumann measurement (VNM) on particle 2 in $\{|+\rangle,|-\rangle\}$ basis given by Eq. (\ref{eq6}) and send 3-c bit message to Bob. Bob makes a suitable unitary transformation (UT) on his qubit 3 to complete the teleportation. On indicating failure, first repeated GBSM may be performed on particles 0 and 2. If success is obtained in this attempt Alice makes VNM on 1 and send the outcomes to Bob, which completes teleportation. If, however, failure is obtained in this attempt also, the second attempt of repeated GBSM may be done. This process of GBSM repetitions on alternate pairs of qubits may continued until perfect teleportation with unit fidelity is attained.}}\label{fig1}
\end{figure}
\section{PQT via 3-qubit NME GHZ state}\label{sec2}
In this scheme, beside the information state, Alice keeps two particles of the 3-qubit NME resource and the third particle of the entangled state belongs to Bob. The schematic diagram of the proposed scheme is shown in Fig.\ref{fig1}. Initially, Alice performs GBSM on the pair of particles consisting of the information state and one of the shared entangled particles. On indicating success, she measures the state of the remaining particle in the eigen basis of $\sigma_x$. She sends her measurement outcomes to Bob through a classical channel. Depending upon the classical input obtained from Alice, Bob performs a suitable unitary operation on his qubit to recover  the information state. However, if GBSM indicates failure then Alice may perform next attempt of repeated GBSM. This time the GBSM carried out on the information qubit and the qubit other than the one used in the last GBSM. The process of GBSM repetitions may be continued until perfect QT with unit fidelity is obtained.
Let Alice has a particle (say, 0) in quantum state 
\begin{equation}
|{I}\rangle_0=a|{0}\rangle_0+b|{1}\rangle_0\label{eq2}
\end{equation} 
where coefficients $a$ and $b$ are unknown to Alice and satisfy the normalization condition, $|a|^2+|b|^2=1$. Suppose, she wish to teleport it to a distant receiver, Bob. To accomplish this task, we consider a 3-qubit NME GHZ state,  
\begin{equation}
|{E}\rangle_{123}=\cos{\chi}|{000}\rangle_{123}+\sin{\chi}|{111}\rangle_{123}\label{eq3}
\end{equation}
with $\chi\;\epsilon\;[0,\pi/4]$, having bipartite entanglement with concurrence ${\cal C}=\sin{2{\chi}}$. Let Alice possesses particles 1 and 2 while particle 3 goes to Bob. We can write the composite state of particles 0, 1, 2 and 3 as,
\begin{eqnarray} 
\nonumber |{\psi}\rangle_{0123} &=&(a|{0}\rangle+b|{1}\rangle)_{0}(\cos{\chi}|{000}\rangle+\sin{\chi}|{111}\rangle)_{123}\\\nonumber
&=& a\cos{\chi}|{0000}\rangle_{0123}+a\sin{\chi}|{0111}\rangle_{0123}\\
&&+b\cos{\chi}|{1000}\rangle_{0123}+b\sin{\chi}|{1111}\rangle_{0123}\label{eq4}.
\end{eqnarray}
The set of complete generalized Bell basis is given by; 
\begin{subequations}
\begin{align}
|{B^{(r,0)}}\rangle=N_r[\cos^r\chi|{00}\rangle+\sin^r\chi|{11}\rangle]\label{eq5.1} \\
|{B^{(r,1)}}\rangle=N_r[\sin^r\chi|{00}\rangle-\cos^r\chi|{11}\rangle]\label{eq5.2} \\
|{B^{(r,2)}}\rangle=N_r[\cos^r\chi|{01}\rangle+\sin^r\chi|{10}\rangle]\label{eqn5.3} \\
|{B^{(r,3)}}\rangle=N_r[\sin^r\chi|{01}\rangle-\cos^r\chi|{10}\rangle]\label{eqn5.4}
\end{align}
\label{eq5}
\end{subequations}\\
where, $N_r=[\cos^{2r}\chi+\sin^{2r}\chi]^{-1/2}$.
Also, the von Neumann measurement (VNM) bases written in computational basis and generalized basis respectively are given by, 
\begin{equation}
|{\pm}\rangle=\frac{1}{\sqrt{2}}[|{0}\rangle\pm|{1}\rangle],\label{eq6}
\end{equation}
\begin{subequations}
\begin{align}
|{C^{(r,0)}}\rangle = N_r[\cos^r\chi|{0}\rangle+\sin^r\chi|{1}\rangle]\label{eq6.1}\\
|{C^{(r,1)}}\rangle = N_r[\sin^r\chi|{0}\rangle-\cos^r\chi|{1}\rangle].\label{eq6.2}
\end{align}
\label{eq6.3}
\end{subequations}
The possible bipartite states and the corresponding Bob's unitary transformation, which will lead to successful Alice's GBSM are given as;
\begin{subequations}
\begin{align}
a|{00}\rangle+b|{11}\rangle=\frac{1}{\sqrt{2}}[|{+}\rangle|{I}\rangle+|{-}\rangle\sigma_z|{I}\rangle]\label{eq7.1} \\
a|{00}\rangle-b|{11}\rangle=\frac{1}{\sqrt{2}}[|{+}\rangle\sigma_z|{I}\rangle+|{-}\rangle|{I}\rangle]\label{eq7.2} \\
a|{11}\rangle+b|{00}\rangle=\frac{1}{\sqrt{2}}[|{+}\rangle\sigma_x|{I}\rangle+|{-}\rangle\sigma_z\sigma_x|{I}\rangle]\label{eq7.3} \\
a|{11}\rangle-b|{00}\rangle=\frac{1}{\sqrt{2}}[|{+}\rangle\sigma_z\sigma_x|{I}\rangle-|{-}\rangle\sigma_x|{I}\rangle]\label{eq7.4} \\
a|{01}\rangle+b|{10}\rangle=\frac{1}{\sqrt{2}}[|{+}\rangle\sigma_x|{I}\rangle+|{-}\rangle\sigma_z\sigma_x|{I}\rangle]\label{eq7.5} \\
a|{01}\rangle-b|{10}\rangle=\frac{1}{\sqrt{2}}[|{+}\rangle\sigma_z\sigma_x|{I}\rangle-|{-}\rangle\sigma_x|{I}\rangle]\label{eq7.6} \\
a|{10}\rangle+b|{01}\rangle=\frac{1}{\sqrt{2}}[|{+}\rangle|{I}\rangle-|{-}\rangle\sigma_z|{I}\rangle]\label{eq7.7} \\
a|{10}\rangle-b|{01}\rangle=\frac{1}{\sqrt{2}}[|{+}\rangle\sigma_z|{I}\rangle-|{-}\rangle\sigma_z|{I}\rangle].\label{eq7.8}
\end{align}
\label{eq7}\\
\end{subequations}
This simply implies that on performing GBSM, if one of the above bipartite states is obtained then Alice makes a von Neumann measurement on her remaining shared entangled particle, in the eigen basis of $\sigma_x$. This will collapses the bipartite state to either $|{I}\rangle$ or one of the three states $\sigma_z|{I}\rangle$, $\sigma_x|{I}\rangle$, $\sigma_x\sigma_z|{I}\rangle$. Alice communicates both her measurement outcomes, which will help Bob to perform a suitable unitary transformation on his particle to recover the information state with unit fidelity. 

\subsection{Original GBSM}\label{subsec1}
Initially, Alice performs GBSM in the orthonormal generalized Bell basis, $|{B^{(1,0)}}\rangle$, $|{B^{(1,1)}}\rangle$, $|{B^{(1,2)}}\rangle$, $|{B^{(1,3)}}\rangle$ on particles 0 and 1. We express Eq.(\ref{eq4}) in this basis and write the input state as,   
\begin{eqnarray}
\nonumber 
|{\psi}\rangle_{0123} &=&|{B^{(1,0)}}\rangle_{01}[a\cos^2\chi|{00}\rangle+b\sin^2\chi|{11}\rangle]_{23}\\\nonumber
&&+|{B^{(1,1)}}\rangle_{01}\cos{\chi}\sin{\chi}[a|{00}\rangle-b|{11}\rangle]_{23} \\\nonumber
&&+|{B^{(1,2)}}\rangle_{01}\cos{\chi}\sin{\chi}[a|{11}\rangle+b|{00}\rangle]_{23}\\
&&+|{B^{(1,3)}}\rangle_{01}[a\sin^2\chi|{11}\rangle-b\cos^2\chi|{00}\rangle]_{23}.\label{eq8}
\end{eqnarray}
The four possible GBSM outcomes  $|{B^{(1,0)}}\rangle$, $|{B^{(1,1)}}\rangle$, $|{B^{(1,2)}}\rangle$, $|{B^{(1,3)}}\rangle$ occurs with corresponding probability,
 \begin{eqnarray}
\nonumber
 P_0&=&|a|^{2}\cos^{4}{\chi}+|b|^{2}\sin^{4}{\chi}\\\nonumber
P_1&=&P_2\nonumber=\sin^{2}{\chi}\cos^{2}{\chi}\\
 P_3&=&|a|^{2}\sin^{4}{\chi}+|b|^{2}\cos^{4}{\chi}\label{eq9}
\end{eqnarray} respectively.\\
 It is clear that GBSM outcomes $|{B^{(1,1)}}\rangle$ and $|{B^{(1,2)}}\rangle$ lead to success. For these cases, Alice measures particle 2 with her in the basis given by Eq.(\ref{eq6}) and communicate the result to Bob. On the other hand, GBSM results $|{B^{(1,0)}}\rangle$ and $|{B^{(1,3)}}\rangle$ will lead to failure, and   Alice go for next attempt of GBSM to improve the success probability.\\ 
 The total success probability in the original GBSM is given by
  \begin{equation}
  P^{(0)}_{success}=2\sin^{2}{\chi}\cos^{2}{\chi}={\cal C}^2/2\label{eq10}
    \end{equation} which is obviously less than unity for NME state (i.e. for ${\cal C}<1$). 
\subsection{First Repeated GBSM}\label{subsec2}
For cases leading to failure, the collapsed states of original GBSM are $|{\psi^{(0)}}\rangle$ and $|{\psi^{(3)}}\rangle$ corresponding to the GBSM results $|{B^{(1,0)}}\rangle$ and $|{B^{(1,3)}}\rangle$ respectively, where, 
 \begin{eqnarray}
\nonumber
|{\psi^{(0)}}\rangle_{0123} &=& |{B^{(1,0)}}\rangle_{01}[a\cos^2\chi|{00}\rangle+b\sin^2\chi|{11}\rangle]_{23}\\\nonumber
&=&[a\cos^3\chi|{0000}\rangle+b\sin^3\chi|{1111}\rangle]_{0123}\\
&&+\cos{\chi}\sin{\chi}(b\sin\chi|{0011}\rangle+a\cos\chi|{1100}\rangle)_{0123}\label{eq11}\\
\nonumber
\end{eqnarray}
and
\begin{eqnarray}
\nonumber|{\psi^{(3)}}\rangle_{0123} &=&|{B^{(1,3)}}\rangle_{01}[a\sin^2\chi|{11}\rangle-b\cos^2\chi|{00}\rangle]_{23}\\\nonumber
&=&[a\sin^3\chi|{0111}\rangle+b\cos^3\chi|{1000}\rangle]_{0123}\\
&&-\cos{\chi}\sin{\chi}(b\cos\chi|{0100}\rangle+a\sin\chi|{1011}\rangle)_{0123}.\label{eq12}
\end{eqnarray} We consider both these cases of failure separately to find out the total success probability that can be obtained when Alice performs the first repeated GBSM on particles 0 and 2.
\paragraph{Case-(1)}
For the collapsed state $|{\psi^{(0)}}\rangle$, Alice makes GBSM on particles 0 and 2 in the orthonormal basis $|{B^{(3,0)}}\rangle$, $|{B^{(3,1)}}\rangle$, $|{B^{(1,2)}}\rangle$, $|{B^{(1,3)}}\rangle$. We expand  $|{\psi^{(0)}}\rangle$ in this basis to obtain 
\begin{eqnarray} 
\nonumber
|{\psi^{(0)}}\rangle_{0123}&=&|{B^{(3,0)}}\rangle_{02}N_3[a\cos^6\chi|{00}\rangle+b\sin^6\chi|{11}\rangle]_{13}\\\nonumber
&&+|{B^{(3,1)}}\rangle_{02}N_3\cos^3{\chi}\sin^3{\chi}[a|{00}\rangle-b|{11}\rangle]_{13}\\\nonumber
&&+|{B^{(1,2)}}\rangle_{02}\cos^3{\chi}\sin^3{\chi}[a|{10}\rangle+b|{01}\rangle]_{13}\\
&&+|{B^{(1,3)}}\rangle_{02}\cos{\chi}\sin{\chi}[b\sin^2\chi|{01}\rangle-a\cos^2\chi|{10}\rangle]_{13}.\label{eq13}
\end{eqnarray}
The four possible GBSM outcomes  $|{B^{(3,0)}}\rangle$, $|{B^{(3,1)}}\rangle$, $|{B^{(1,2)}}\rangle$, $|{B^{(1,3)}}\rangle$ occurs with corresponding probability,  

\begin{eqnarray}
\nonumber
&&P_{00}=N_3^2(|a|^{2}\cos^{12}{\chi}+|b|^{2}\sin^{12}{\chi}),\\\nonumber
&&P_{01}=N_3^2\cos^{6}{\chi}\sin^{6}{\chi}\\\nonumber
&&P_{02}=\cos^{4}{\chi}\sin^{4}{\chi}\\
&&P_{03}=\cos^{2}{\chi}\sin^{2}{\chi}(|a|^{2}\cos^{4}{\chi}+|b|^{2}\sin^{4}{\chi}).\label{eq14}
 \end{eqnarray}
Clearly, the GBSM results $|{B^{(3,1)}}\rangle$ and $|{B^{(1,2)}}\rangle$ lead to success while GBSM result $|{B^{(3,0)}}\rangle$ or $|{B^{(1,3)}}\rangle$ gives failure. For the case when failure is obtained, Alice performs next repeated GBSM on particles 0 and 1.\\  
 The increment in the success probability after first repeated GBSM on $|\psi^{(0)}\rangle$ is given by,
\begin{equation}
P_{01}+P_{02}=\frac{cos^{6}{\chi}\sin^{6}{\chi}}{cos^{6}{\chi}+\sin^{6}{\chi}}+cos^{4}{\chi}\sin^{4}{\chi}.\label{eq15}
  \end{equation}
\paragraph{Case-(2)} For the collapsed state $|{\psi^{(3)}}\rangle$, for perfect QT, Alice performs GBSM on particles 0 and 2 in the Bell basis $|{B^{(1,0)}}\rangle$, $|{B^{(1,1)}}\rangle$, $|{B^{(3,2)}}\rangle$, $|{B^{(3,3)}}\rangle$. We expand $|{\psi^{(3)}}\rangle$ in this basis as,
\begin{eqnarray}
\nonumber
|{\psi^{(3)}}\rangle_{0123}&=&|{B^{(1,0)}}\rangle_{02}\cos{\chi}\sin{\chi}[a\sin^2\chi|{01}\rangle+b\cos^2\chi|{10}\rangle]_{13}\\\nonumber
&&+|{B^{(1,1)}}\rangle_{02}\cos^2{\chi}\sin^2{\chi}[-a|{01}\rangle+b|{10}\rangle]_{13}\\\nonumber
&&+|{B^{(3,2)}}\rangle_{02}N_3\cos^3{\chi}\sin^3{\chi}[a|{11}\rangle+b|{00}\rangle]_{13}\\
&&+|{B^{(3,3)}}\rangle_{02}N_3[a\sin^6{\chi}|{11}\rangle-b\cos^6{\chi}|{00}\rangle]_{13}.\label{eq16}
\end{eqnarray}
The four possible GBSM outcomes $|{B^{(1,0)}}\rangle$, $|{B^{(1,1)}}\rangle$, $|{B^{(3,2)}}\rangle$, $|{B^{(3,3)}}\rangle$ occurs with corresponding probability,  
\begin{eqnarray}
\nonumber
&&P_{30}=\cos^{2}{\chi}\sin^{2}{\chi}(|a|^{2}\sin^{4}{\chi}+|b|^{2}\cos^{4}{\chi})\\\nonumber
&&P_{31}=\cos^{4}{\chi}\sin^{4}{\chi}\\\nonumber
&&P_{32}=N_3^2\cos^{6}{\chi}\sin^{6}{\chi}\\
&&P_{33}=N_3^2(|a|^{2}\cos^{12}{\chi}+|b|^{2}\sin^{12}{\chi})\label{eq17}
 \end{eqnarray}
 respectively.\\
 It is clear that the GBSM results $|{B^{(1,1)}}\rangle$ and $|{B^{(3,2)}}\rangle$ lead to success while GBSM results $|{B^{(1,0)}}\rangle$ and $|{B^{(3,3)}}\rangle$ give failure. For the latter case, Alice repeats GBSM.\\ 
The additional installment of success probability after first repeated GBSM on $|\psi^{(0)}\rangle$ is given by,
\begin{equation}
P_{31}+P_{32}=\frac{cos^{6}{\chi}\sin^{6}{\chi}}{cos^{6}{\chi}+\sin^{6}{\chi}}+cos^{4}{\chi}\sin^{4}{\chi}.\label{eq18}
\end{equation} 
It is evident that the total success probability increases on performing first repeated GBSM on the collapsed state. The overall success probability after the first repeated GBSM is given by, 
\begin{eqnarray}
\nonumber
{\cal P}^{(1)}_{success} &=& {\cal P}^{(0)}_{success}+P_{01}+P_{02}+P_{31}+P_{32}\\\nonumber
&=&\frac{2 cos^{6}{\chi}\sin^{6}{\chi}}{cos^{6}{\chi}+\sin^{6}{\chi}}+2cos^{4}{\chi}\sin^{4}{\chi}+2 cos^{2}{\chi}\sin^{2}{\chi}\\
&=&{\cal P}^{(0)}_{success}+\frac{{\cal C}^{4}}{8}+\frac{{\cal C}^{6}}{8(4-3{\cal C}^{2})}.\label{eq19}
\end{eqnarray}
 \subsection{Second Repeated GBSM} \label{subsec3}
Only four out of eight mutually exclusive GBSM outcomes were successful in the first repeated attempt. The rest of the cases lead to failure, prompting a second repeated attempt of GBSM on particles 0 and 1. The two failure cases of first repeated GBSM on $|{\psi^{(0)}}\rangle$ are $|{\psi^{(00)}}\rangle$ and $|{\psi^{(03)}}\rangle$ corresponding to the GBSM results $|{B^{(3,0)}}\rangle$ and $|{B^{(1,3)}}\rangle$, respectively, where  
\begin{eqnarray}
\nonumber
|{\psi^{(00)}}\rangle_{0123} &=&|{B^{(3,0)}}\rangle_{02}N_3^2[a\cos^6\chi|{00}\rangle+b\sin^6\chi|{11}\rangle]_{13}\\\nonumber
&=&N_3^2[a\cos^9\chi|{0000}\rangle+b\sin^9\chi|{1111}\rangle]_{0123}\\
&&+N_3^2\cos^3{\chi}\sin^3{\chi}(a\cos^3\chi|{1010}\rangle+b\sin^3\chi|{0101}\rangle)_{0123},\label{eq20}\\
\nonumber
|{\psi^{(03)}}\rangle_{0123} &=&|{B^{(1,3)}}\rangle_{02}\cos{\chi}\sin{\chi}[b\sin^2\chi|{01}\rangle-a\cos^2\chi|{10}\rangle]_{13}\\\nonumber
&=&\cos{\chi}\sin{\chi}[a\cos^3\chi|{1100}\rangle+b\sin^3\chi|{0011}\rangle]_{0123}\\
&&-\cos^2{\chi}\sin^2{\chi}(a\cos\chi|{0110}\rangle+b\sin\chi|{1001}\rangle)_{0123}.\label{eq21}
\end{eqnarray}
Similarly, the two failure cases of first repeated GBSM on $|{\psi^{(3)}}\rangle$ are $|{\psi^{(30)}}\rangle$ and $|{\psi^{(33)}}\rangle$ corresponding to the GBSM results $|{B^{(1,0)}}\rangle$ and $|{B^{(3,3)}}\rangle$ respectively, where,
\begin{eqnarray}
\nonumber|{\psi^{(30)}}\rangle_{0123} &=&|{B^{(1,0)}}\rangle_{02}\cos{\chi}\sin{\chi}[a\sin^2\chi|{01}\rangle+b\cos^2\chi|{10}\rangle]_{13}\\\nonumber
&=&\cos{\chi}\sin{\chi}[a\sin^3\chi|{1011}\rangle+b\cos^3\chi|{0100}\rangle]_{0123}\\
&&+\cos^2{\chi}\sin^2{\chi}(a\sin\chi|{0001}\rangle+b\cos\chi|{1110}\rangle)_{0123},\label{eq22}\\
\nonumber
|{\psi^{(33)}}\rangle_{0123} &=&|{B^{(3,3)}}\rangle_{02}N_3^2[a\sin^6{\chi}|{11}\rangle-b\cos^6{\chi}|{00}\rangle]_{13}\\\nonumber
&=&N_3^2[a\sin^9\chi|{0111}\rangle+b\cos^9\chi|{1000}\rangle]_{0123}\\
&&-N_3^2\cos^{3}{\chi}\sin^{3}{\chi}(a\sin^{3}{\chi}|{1101}\rangle+b\cos^{3}{\chi}|{0010}\rangle)_{0123}.\label{eq23}
\end{eqnarray}
The choice of Bell bases, possible GBSM outcomes and the corresponding probability of occurrence for each of the above four cases are summarized in Table~\ref{tab1}, which clearly shows that,\\
\begin{table}[h!]
\caption{Alice GBSM result in second repeated GBSM on particles 0 and 1}\label{tab1}
\resizebox{\linewidth}{!}{\begin{tabular}{l l l p{7cm}  c }\\
\hline\\
Input state & GBSM  & Unnormalized state of qubits 2 and 3  & Probability \newline $P_{ijk}$ for $|\psi^{(ij)}\rangle$ and $|{B^{(r,k)}}\rangle$  & Fidelity\\[0.5ex]
\hline\\
$|\psi^{(00)}\rangle$ & $|{B^{(9,0)}}\rangle$ & $N_3^2 N_9[a\cos^{18}{\chi}|{00}\rangle+b\sin^{18}{\chi}|{11}\rangle]_{23}$ & $N_3^4 N_9^2[|a|^{2}\cos^{36}{\chi}+|b|^{2}\sin^{36}{\chi}]$  & $\neq 1$\\
& $|{B^{(9,1)}}\rangle$ & $N_3^2 N_9\cos^{9}{\chi}\sin^{9}{\chi}[a|{00}\rangle-b|{11}\rangle]_{23}$ & $N_3^4 N_{9}^{2}[\cos^{18}{\chi}\sin^{18}{\chi}]$ & $1$\\
& $|{B^{(3,2)}}\rangle$ & $N_3^3\cos^{6}{\chi}\sin^{6}{\chi}[a|{10}\rangle+b|{01}\rangle]_{23}$ & $N_3^6[\cos^{12}{\chi}\sin^{12}{\chi}]$ & $1$\\
& $|{B^{(3,3)}}\rangle$ & $N_3^3\cos^{3}{\chi}\sin^{3}{\chi}[-a\cos^{6}{\chi}|{10}\rangle+b\sin^{6}{\chi}|{01}\rangle]_{23}$ &  $N_{3}^6\cos^{6}{\chi}\sin^{6}{\chi}[|a|^{2}\cos^{12}{\chi}+|b|^{2}\sin^{12}{\chi}]$ & $\neq 1$\\[0.5ex]
\hline\\
$|\psi^{(03)}\rangle$ & $|{B^{(3,0)}}\rangle$ & $ N_3\cos^{4}{\chi}\sin^{4}{\chi}[a|{00}\rangle+b|{11}\rangle]_{23}$ & $N_{3}^{2}[\cos^{8}{\chi}\sin^{8}{\chi}]$ & $1$\\
& $|{B^{(3,1)}}\rangle$ & $ N_3\cos{\chi}\sin{\chi}[-a\cos^{6}{\chi}|{00}\rangle+b\sin^{6}{\chi}|{11}\rangle]_{23}$ &  $N_3^2\cos^{2}{\chi}\sin^{2}{\chi}[|a|^{2}\cos^{12}{\chi}+|b|^{2}\sin^{12}{\chi}]$ & $\neq 1$\\
& $|{B^{(1,2)}}\rangle$ & $ \cos^2{\chi}\sin^2{\chi}[a\cos^{2}{\chi}|{10}\rangle+b\sin^{2}{\chi}|{01}\rangle]_{23}$ &  $\cos^{4}{\chi}\sin^{4}{\chi}[|a|^{2}\cos^{4}{\chi}+|b|^{2}\sin^{4}{\chi}]$ & $\neq 1$\\
& $|{B^{(1,3)}}\rangle$ & $ \cos^{3}{\chi}\sin^{3}{\chi}[a|{10}\rangle-b|{01}\rangle]_{23}$ &  $\cos^{6}{\chi}\sin^{6}{\chi}$ & $1$\\[0.5ex]
\hline\\
$|\psi^{(30)}\rangle$ & $|{B^{(1,0)}}\rangle$ & $\cos^{3}{\chi}\sin^{3}{\chi}[a|{01}\rangle+b|{10}\rangle]_{23}$ & $\cos^{6}{\chi}\sin^{6}{\chi}$  & $ 1$\\
& $|{B^{(1,1)}}\rangle$ & $\cos^2{\chi}\sin^2{\chi}[a\sin^{2}{\chi}|{01}\rangle-b\cos^{2}{\chi}|{10}\rangle]_{23}$ &  $\cos^{4}{\chi}\sin^{4}{\chi}[|a|^{2}\sin^{4}{\chi}+|b|^{2}\cos^{4}{\chi}]$ & $\neq 1$\\
& $|{B^{(3,2)}}\rangle$ & $N_3\cos{\chi}\sin{\chi}[b\cos^{6}{\chi}|{00}\rangle+a\sin^{6}{\chi}|{11}\rangle]_{23}$ &  $N_{3}^{2}\cos^2{\chi}\sin^2{\chi}[|a|^{2}\sin^{12}{\chi}+|b|^{2}\cos^{12}{\chi}]$ & $\neq 1$\\
& $|{B^{(3,3)}}\rangle$ & $N_3\cos^{4}{\chi}\sin^{4}{\chi}[b|{00}\rangle-a|{11}\rangle]_{23}$ &  $N_{3}^{2}\cos^{8}{\chi}\sin^{8}{\chi}$ & $1$\\[0.5ex]
\hline\\
$|\psi^{(33)}\rangle$ & $|{B^{(3,0)}}\rangle$ & $N_3^3\cos^{3}{\chi}\sin^{3}{\chi}[a\sin^{6}{\chi}|{01}\rangle+b\cos^{6}{\chi}|{10}\rangle]_{23}$ & $N_{3}^6\cos^{6}{\chi}\sin^{6}{\chi}[|a|^{2}\sin^{12}{\chi}+|b|^{2}\cos^{12}{\chi}]$ & $\neq 1$\\
& $|{B^{(3,1)}}\rangle$ & $ N_3^3\cos^{6}{\chi}\sin^{6}{\chi}[a|{01}\rangle-b|{10}\rangle]_{23}$ & $N_3^6[\cos^{12}{\chi}\sin^{12}{\chi}]$ & $1$\\
& $|{B^{(9,2)}}\rangle$ & $N_3^2 N_9\cos^{9}{\chi}\sin^{9}{\chi}[a|{11}\rangle+b|{00}\rangle]_{23}$ & $N_3^4 N_{9}^{2}[\cos^{18}{\chi}\sin^{18}{\chi}]$ & $1$\\
& $|{B^{(9,3)}}\rangle$ & $N_3^2 N_9[a\sin^{18}{\chi}|{11}\rangle-b\cos^{18}{\chi}|{00}\rangle]_{23}$ & $N_3^4 N_9^2[|a|^{2}\sin^{36}{\chi}+|b|^{2}\cos^{36}{\chi}]$ & $\neq 1$\\[0.5ex]
\hline
\end{tabular}}
\end{table}\\
(1) For the collapsed state $|{\psi^{(00)}}\rangle$, Alice performs GBSM in the orthonormal generalized Bell basis, $|{B^{(9,0)}}\rangle$, $|{B^{(9,1)}}\rangle$, $|{B^{(3,2)}}\rangle$, $|{B^{(3,3)}}\rangle$. QT is successful for the GBSM result $|{B^{(9,1)}}\rangle$ or $|{B^{(3,2)}}\rangle$ whereas for the GBSM outcome $|{B^{(9,0)}}\rangle$ or $|{B^{(3,3)}}\rangle$ failure is indicated. The additional success probability due to the second repeated GBSM on $|{\psi^{(00)}}\rangle$ is $P_{001}+P_{002}$.\\\\
(2) For the collapsed state $|{\psi^{(03)}}\rangle_{0123}$, Alice chooses the orthonormal generalized Bell basis as $|{B^{(3,0)}}\rangle$, $|{B^{(3,1)}}\rangle$, $|{B^{(1,2)}}\rangle$, $|{B^{(1,3)}}\rangle$ and performs second repeated GBSM. We see that QT is successful for the GBSM result $|{B^{(3,0)}}\rangle$ or $|{B^{(1,3)}}\rangle$, whereas for GBSM outcome $|{B^{(3,1)}}\rangle$ or $|{B^{(1,2)}}\rangle$ failure is obtained. The additional success probability due to the second repeated GBSM on $|{\psi^{(03)}}\rangle$ is $P_{030}+P_{033}$.\\\\
(3) For the collapsed state $|{\psi^{(30)}}\rangle_{0123}$, Alice makes GBSM in the orthonormal generalized Bell basis, $|{B^{(1,0)}}\rangle$, $|{B^{(1,1)}}\rangle$, $|{B^{(3,2)}}\rangle$, $|{B^{(3,3)}}\rangle$. For this case, QT is successful for the GBSM result $|{B^{(1,0)}}\rangle$ or $|{B^{(3,3)}}\rangle$, whereas for the GBSM outcome $|{B^{(1,1)}}\rangle$ or $|{B^{(3,2)}}\rangle$ failure is obtained. The additional success probability due to this second repeated GBSM on $|{\psi^{(30)}}\rangle$ is $P_{300}+P_{303}$.\\\\
(4) For the collapsed state $|{\psi^{(33)}}\rangle_{0123}$, Alice chooses the orthonormal generalized Bell basis, $|{B^{(3,0)}}\rangle$, $|{B^{(3,1)}}\rangle$, $|{B^{(9,2)}}\rangle$, $|{B^{(9,3)}}\rangle$ and performs second repeated GBSM. QT is successful for the GBSM result $|{B^{(3,1)}}\rangle$ or $|{B^{(9,2)}}\rangle$, whereas for the GBSM result $|{B^{(3,0)}}\rangle$ or $|{B^{(9,3)}}\rangle$ failure is obtained. Therefore, the additional success due to this second repeated GBSM on $|{\psi^{(33)}}\rangle$ is $P_{331}+P_{332}$.

Therefore, the total success probability at the end of second repeated GBSM is given by,
\begin{equation}
{\cal P}^{(2)}_{success}={\cal P}^{(1)}_{success} +P_{001}+P_{002}+P_{030}+P_{033}+P_{300}+P_{303}+P_{331}+P_{332}.\label{eq24}
\end{equation}
On substituting the values from Table~\ref{tab1} and putting, $\sin{2\chi}={\cal C}$, we get,
\begin{equation}
{\cal P}^{(2)}_{success}={\cal P}^{(1)}_{success}+\frac{{\cal C}^{6}}{32}+\frac{{\cal C}^{8}}{32(4-3{\cal C}^{2})}+\frac{{\cal C}^{12}}{32(4-3{\cal C}^{2})^{3}}+\frac{{\cal C}^{18}}{32(4-3{\cal C}^{2})^{3}[4(4-3{\cal C}^{2})^{2}-3{\cal C}^{6}]}.\label{eq25}
\end{equation}

\subsection{Third Repeated GBSM} \label{subsec4}
A failure obtained in the second attempt shall demand for a third repeated attempt of GBSM. This will involve eight cases of failure that are obtained in the second attempt of GBSM. Following the procedure as before, i.e., choosing a suitable generalized Bell basis for performing GBSM, and calculating success probability for each cases and adding them will lead to total success probability at the end of this attempt as,
\begin{eqnarray}
\nonumber
{\cal P}^{(3)}_{Success} &=&{\cal P}^{(2)}_{success}+\frac{{\cal C}^{8}}{128}+\frac{{\cal C}^{10}}{128(4-3{\cal C}^{2})}+\frac{{\cal C}^{14}}{128(4-3{\cal C}^{2})^{3}}+\frac{{\cal C}^{18}}{128(4-3{\cal C}^{2})^{5}}\\\nonumber
&&+\frac{{\cal C}^{54}}{128(4-3{\cal C}^{2})^{5}[4(4-3{\cal C}^{2})^{2}-3{\cal C}^{6}]^{3}[4(4-3{\cal C}^{2})^{2}((4-3{\cal C}^{2})^{2}-{\cal C}^{6})^{2}-{\cal C}^{18}]}\\\nonumber
&&+\frac{{\cal C}^{24}}{128(4-3{\cal C}^{2})^{5}[4(4-3{\cal C}^{2})^{2}-3{\cal C}^{6}]}+\frac{{\cal C}^{36}}{128(4-3{\cal C}^{2})^{5}[4(4-3{\cal C}^{2})^{2}-3{\cal C}^{6}]^{3}}\\
&&+\frac{{\cal C}^{20}}{128(4-3{\cal C}^{2})^{3}[4(4-3{\cal C}^{2})^{2}-3{\cal C}^{6}]}\label{eq26}
\end{eqnarray}
\section{Results and Discussions}\label{sec3}
\begin{figure}[h!]
\centering
\begin{subfigure}{0.48\textwidth}
    \includegraphics[width=\textwidth]{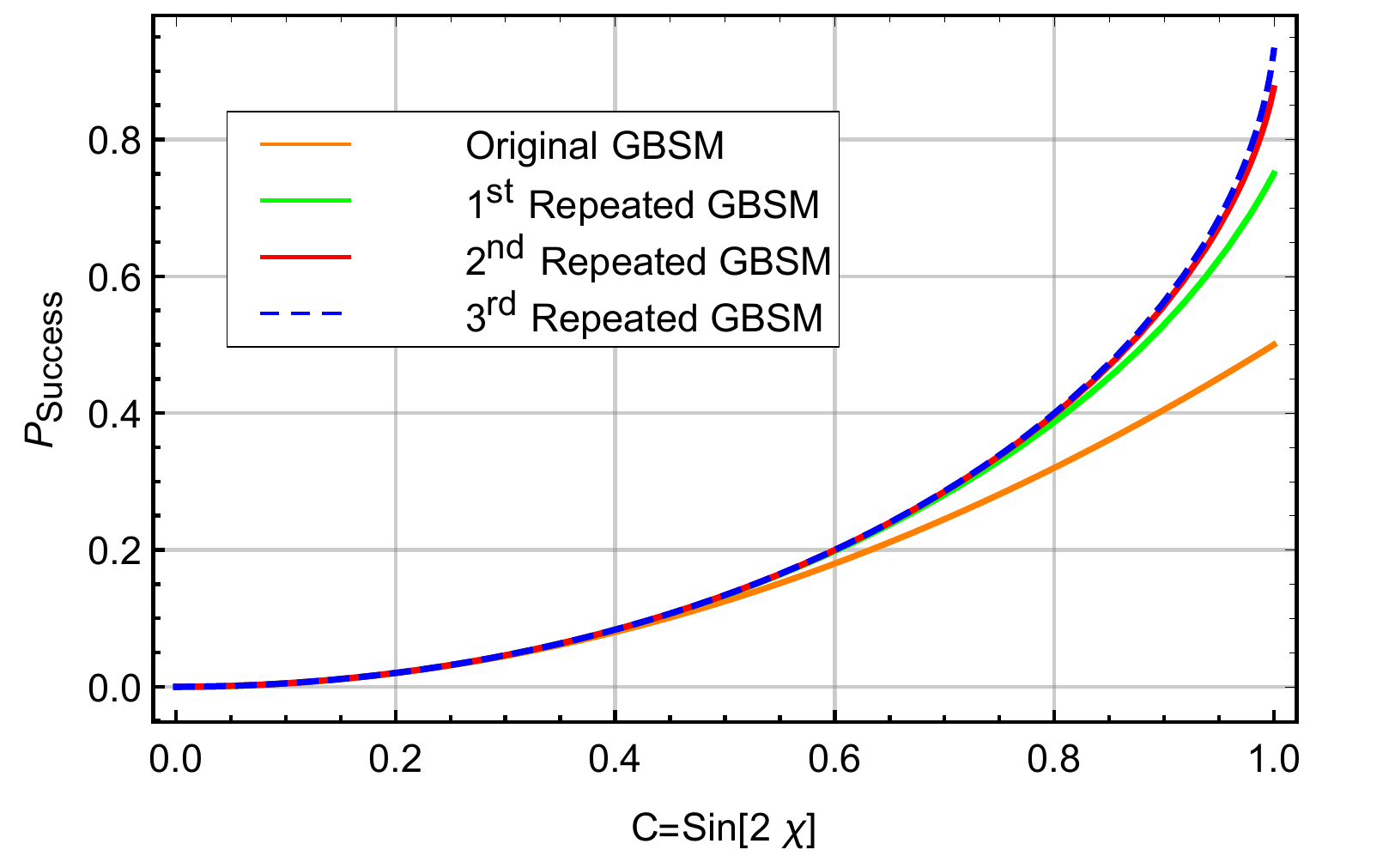}
    \caption{}
    \label{fig2}
\end{subfigure}
\hfill
\begin{subfigure}{0.48\textwidth}
    \includegraphics[width=\textwidth]{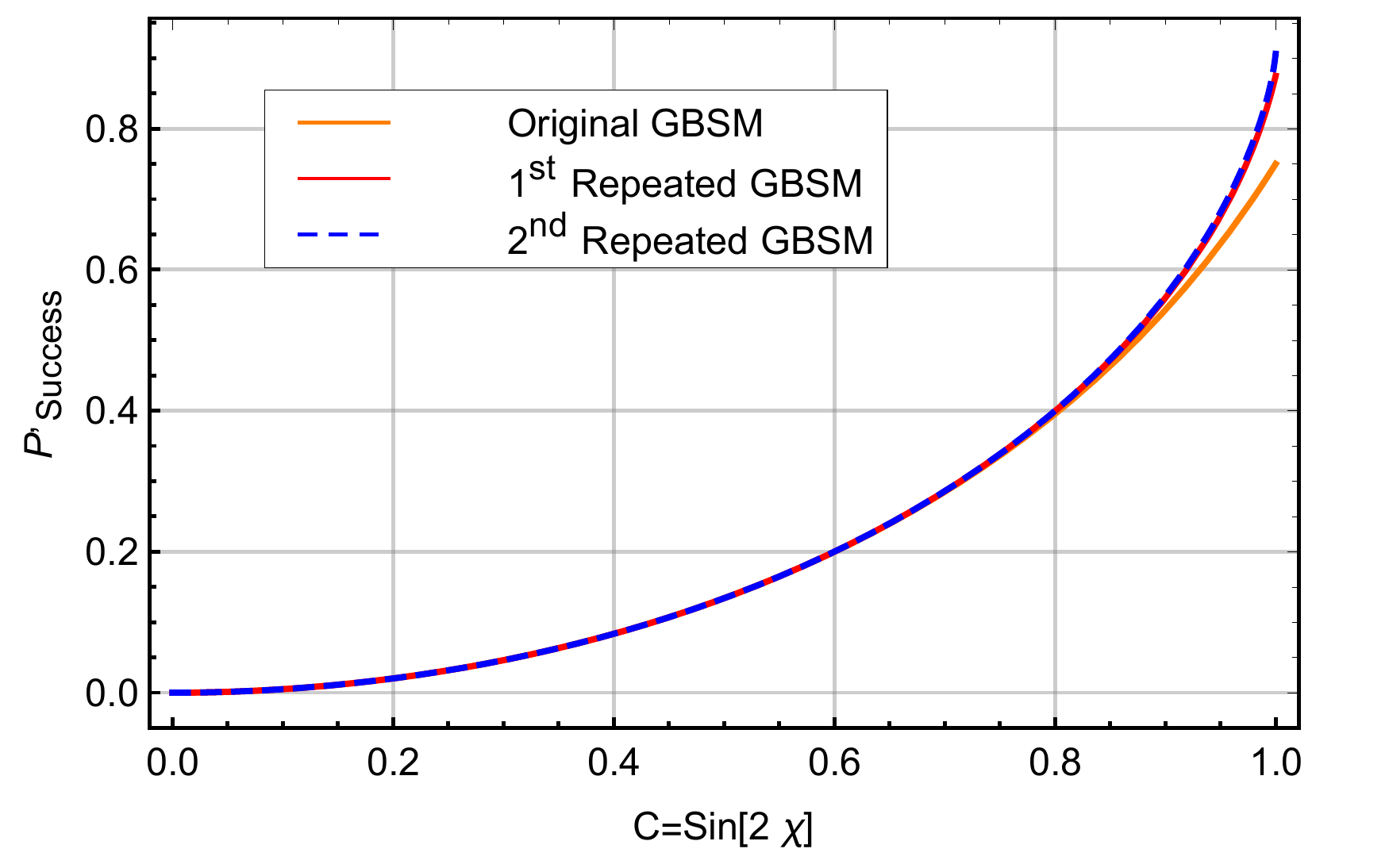}
    \caption{}
    \label{fig3}
\end{subfigure}
        
\caption{Variation of success probability with the bipartite concurrence of the 3-qubit NME GHZ state, (a) with continuation of repeated GBSM (b) with suitably chosen VNM in final attempt.}
\label{fig23}
\end{figure}

The variation of success probability with the bipartite concurrence of the 3-qubit NME GHZ resource state has been plotted in Fig. \ref{fig2}. It is evident that for NME resource which have nearly unit bipartite concurrence (i.e. ${\cal C}\approx 1$) the success probability increases with number of GBSM repetitions. The net increase in the success probability in going from \textit{n} to \textit{n+1} repetition, however, decreases as \textit{n} increases.

\begin{figure}[h!]
\centering
\begin{subfigure}{0.48\textwidth}
    \includegraphics[width=\textwidth]{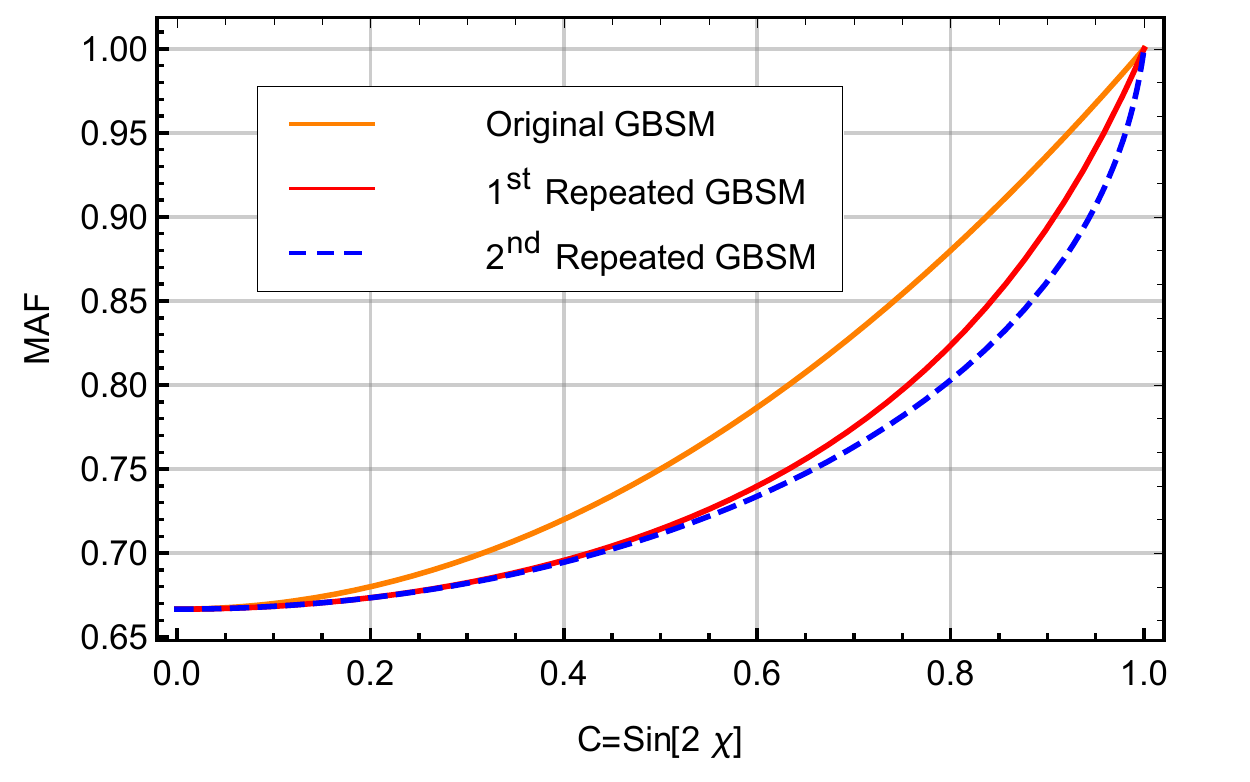}
    \caption{}
    \label{fig4}
\end{subfigure}
\hfill
\begin{subfigure}{0.48\textwidth}
    \includegraphics[width=\textwidth]{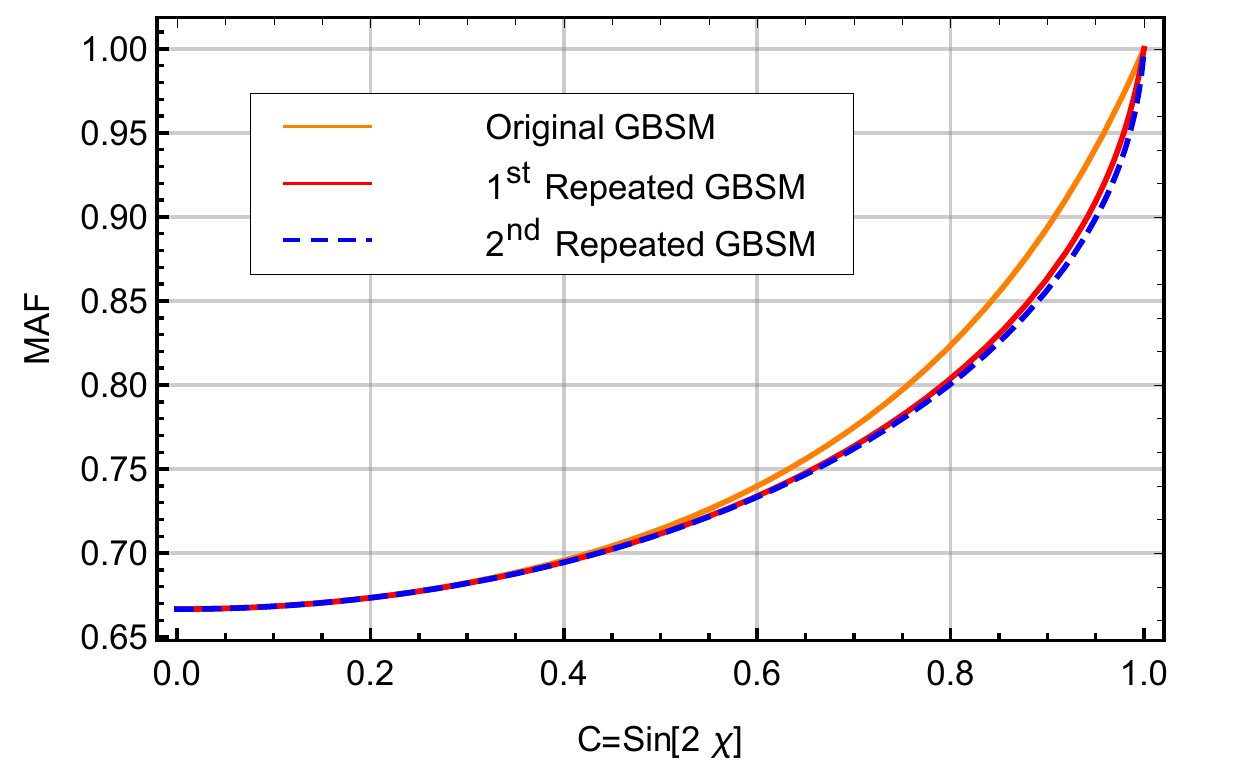}
    \caption{}
    \label{fig5}
\end{subfigure}
        
\caption{Variation of maximal average fidelity with the bipartite concurrence of the 3-qubit NME GHZ state, (a) with VNM in final attempt (b) with a suitbly chosen VNM in final attempt.}
\label{fig45}
\end{figure}

Theoretically, the process of repeating GBSM on alternate pairs of qubits may be continued until perfect QT with unit fidelity is achieved. However, in a more realistic scenario, the number of GBSM repetitions may be limited to, (say) $n$. Let us consider a situation  where after performing GBSM \textit{n} times, failure is obtained, and Alice wish to send the information to Bob. In that case, Alice has two options: either she measures her single (remaining) particle (1) in the eigen basis of $\sigma_x$, or (2) in a suitably chosen generalized VNM basis given by Eq.(\ref{eq6.3}). It is seen that for the latter case, the success probability can be further improved. The improved success probability up to the second repeated GBSM have been plotted with respect to the bipartite concurrence in Fig. \ref{fig3}. Also the variation of maximal average fidelity (MAF) with respect to the bipartite concurrence for both these cases, has been shown in Fig. \ref{fig45}. We see that in both the above cases, although, MAF is decreasing with an increase in the GBSM repetitions, for the latter case i.e. for the generalized VNM by Alice, the success probability can be further improved.
\section{Conclusions}\label{sec4}

It is worthwhile to compare our present scheme with the previous scheme of Javed et al. \cite{javed2021high} for transmission of a single qubit using a 2-qubit NME resource. Javed et al. \cite{javed2021high} scheme has two major drawbacks: Firstly, the transmission of information state from Alice to Bob is delayed, since, after performing GBSM Alice has to send the particle and a 2 c-bit message to Bob. Secondly, the quantum channel used by Alice to transmit the particle to Bob will inevitably introduce losses in the information state, thereby, reducing the fidelity of teleportation.

Our present scheme is capable of circumventing both these issues. Apart from initially distributing the 3-qubit NME resource, no further need arises to send the qubit over a quantum channel. Thus, the scheme is essentially a PQT. This apparently solves both the issues viz. the delayed information transmission and the information loss due to interaction with the transmission channel. In our scheme,  Alice performs GBSM collectively on the information state and one of her qubits of the NME resource. On indicating success, she performs a projective measurement on her remaining qubit (apart from the one used in last GBSM) in the eigen basis of $\sigma_x$  and communicates both of her measurement results to Bob through a classical channel. This helps Bob to reconstruct the information state and complete the teleportation. Therefore, our present scheme of PQT is more obvious and experimentally feasible compared to the one reported in \cite{javed2021high}.   	 
\section*{Acknowledgement}\label{sec5}
SJ, RKP, PSY dedicate this paper in the memory of Prof. Ranjana Prakash and Prof. Hari Prakash. May their soul rest in peace. We are thankful to Prof. N Chandra, Dr. V Verma and Dr. O N Verma for valuable suggestions. SJ is thankful to UGC scheme of Maulana Azad National Fellowship for providing financial support. RKP is thankful to UGC for providing financial assistance under CSIR-UGC SRF fellowship.    

\bibliographystyle{unsrt} 
\bibliography{cqt}

\end{document}